\documentclass[sigconf]{acmart}

\AtBeginDocument{%
  }

\copyrightyear{2026}
\acmYear{2026}
\setcopyright{cc}
\setcctype{by}
\acmConference[CHI EA '26]{Extended Abstracts of the 2026 CHI Conference on Human Factors in Computing Systems}{April 13--17, 2026}{Barcelona, Spain}
\acmBooktitle{Extended Abstracts of the 2026 CHI Conference on Human Factors in Computing Systems (CHI EA '26), April 13--17, 2026, Barcelona, Spain}
\acmDOI{10.1145/3772363.3799105}
\acmISBN{979-8-4007-2281-3/2026/04}
\acmISBN{978-1-4503-XXXX-X/2018/06}

\begin{document}

\title{The Perceptual Gap: Why We Need Accessible XAI for Assistive Technologies}

\author{Shadab H. Choudhury}
\email{shadabc1@umbc.edu}
\affiliation{%
  \institution{University of Maryland, Baltimore County}
  \city{Baltimore}
  \state{Maryland}
  \country{USA}
}

\renewcommand{\shortauthors}{Choudhury et al.}

\begin{abstract}
Artificial intelligence systems are widely used by people with sensory disabilities, like loss of vision or hearing, to help perceive or navigate the world around them. This includes tasks like describing an image or object they cannot touch, reading documents, automatically captioning speech, and so on. Presently, models used for these tasks are based on deep neural networks and are thusly black boxes. Explainable AI (XAI) describes methods that can explain why a model gave the output it did. However, existing XAI methodologies are rarely accessible or designed with disabled users in mind. In this paper, we survey existing work in XAI with a focus on human-centered and accessibility-centered approaches or evaluations. We show that there is next-to-no XAI work that accounts for people with sensory disabilities, that many typical explanations are difficult for them to comprehend, and propose possible avenues for future work in Accessible Human-Centered XAI.
\end{abstract}


\begin{teaserfigure}
  \includegraphics[width=\textwidth]{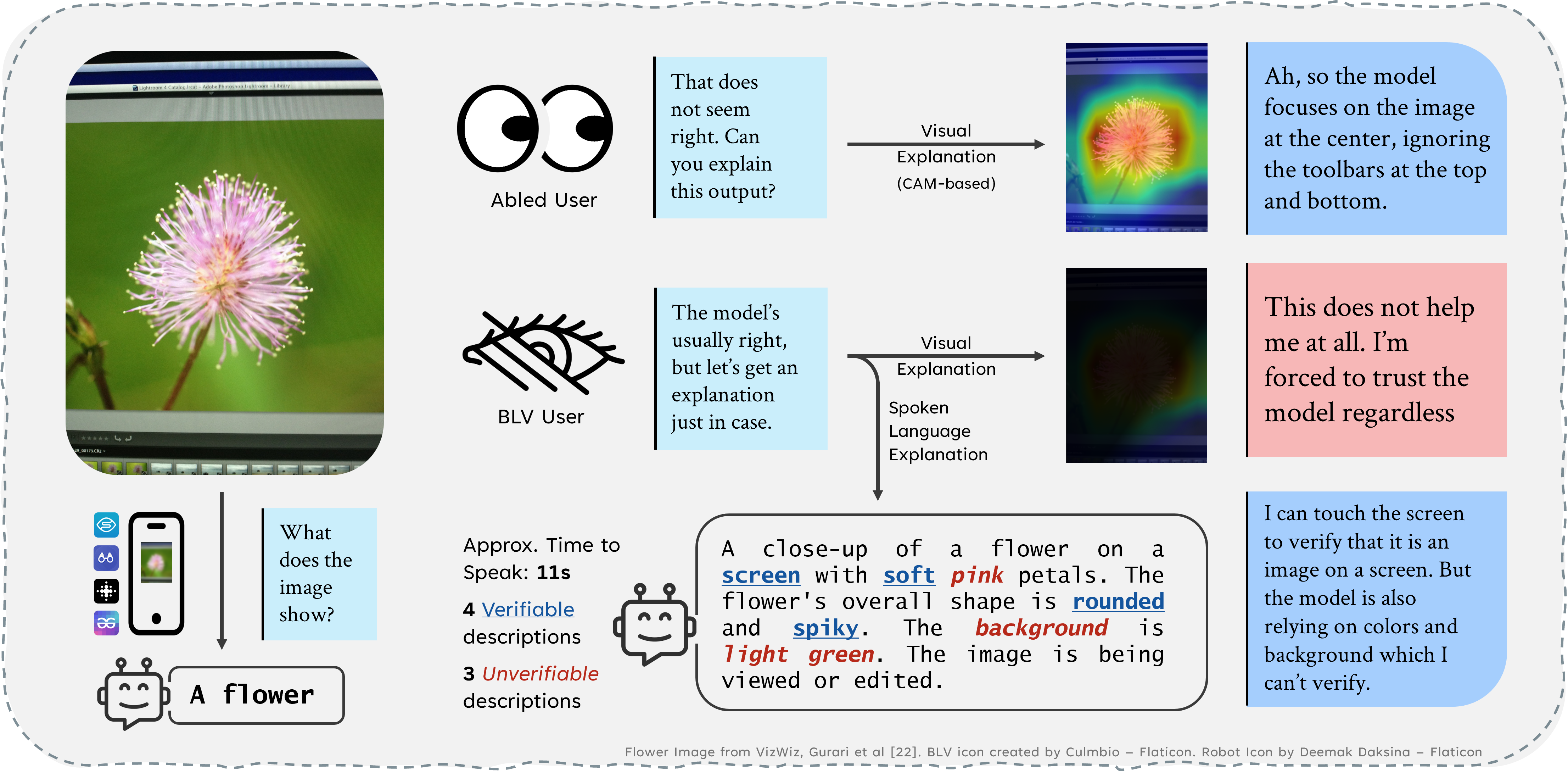}
  \caption{A depiction of the process of using a image recognition app, how visual explainability methods are flawed for the primary target userbase of such applications, and how verifiability remains a problem with language descriptions.}
  \Description{An image of a flower on a computer screen, loaded into an image processing software with toolbars visible at the top and bottom. The user asks, "What is this image showing?". The app responds with "A flower. A sighted user says "That does not seem right. Can you explain this output?" and based on the explanation, deduces that the model does not recognize the toolbars at the top and bottom, only the flower image in the middle. A Blind or Low-Vision user may think, at this stage, "The model’s usually right, but let’s get an explanation just in case." However, the explanation is a visual heatmap they are not able to see (nor you, if you are in the above group and using a screenreader). The alternative explanation is a language description stating 'A close-up of a flower on a screen with soft pink petals. The flower's overall shape is rounded and spiky. The background is light green. The image is being viewed or edited.' This description takes about 11 seconds to speak out loud and contains four verifiable descriptions (screen, soft, rounded, spiky) and three unverifiable descriptions (pink, background, light green). The Blind or Low-Vision user now thinks, "I can touch the screen to verify that it is an image on a screen. But the model is also relying on colors and background which I can't verify."}
  \label{fig:VisualExplanationHeader}
\end{teaserfigure}

\maketitle

\section{Introduction}
\label{sec:Introduction}

Using Artificial Intelligence (AI), digital systems can now emulate human behavior under a strict, narrow scope. For example, an object recognition model, given an image of an object, can return the name or description of that object, just like a person looking at the picture could. Assistive devices and software designed today regularly take advantage of these capabilities. An individual who is blind or has low-vision (BLV) could use such a model to ``see'' nearby objects without having to touch and feel them. Apps such as \textit{Seeing AI} \footnote{https://www.seeingai.com/}, \textit{Lookout} \footnote{https://play.google.com/store/apps/details?id=com.google.android.apps.accessibility.\\reveal}, \textit{Envision} \footnote{https://www.letsenvision.com/}, and \textit{TapTapSee} \footnote{https://taptapseeapp.com/} all use AI. These are not niche applications either --- Lookout and TapTapSee both have over 500k downloads on Android devices, while the other two have over 100k downloads.

Similarly, an individual who is deaf or hard-of-hearing (DHH) can use a speech captioning model to convert spoken language that they can't hear to written language that they can read. \textit{Otter} \footnote{https://otter.ai/} is the most widely used standalone app for this (with north of 5 million downloads on Android), but most modern devices and video calling software now also have built-in closed captioning. 


How does one know if the output of a model is accurate or reliable? One way is to have an intimate knowledge of how the model was designed and trained. The other way is to study the way the model processes an input and arrives to a particular conclusion. 'Interpretability' and 'Explainability' both cover this from different perspectives --- Interpretability focuses on understanding how the model works in a way that machine learning practitioners can use the insights \cite{kaur_interpreting_2020}. Explainability attempts to describe or exhibit the model's behaviour in such a way that a layperson can understand how the model works to some extent \cite{dwivedi_explainable_2023}. Explainability is generally \textit{post-hoc}, which means there is no need to dive into a model's internals or design processes to understand it.

Explainability methods like SHAP \cite{lundberg_unified_2017} or LIME \cite{ribeiro_why_2016} explain the outputs by listing how important the different components of the input (a.k.a. features) were for the final output. However, this is only feasible for inputs with few features, such as a table of values. CAM or GRAD-CAM \cite{zhou_learning_2016, selvaraju_grad-cam_2017} are more useful for the object recognition models typically used in apps for BLV users; they generate a 'heat map' overlaid on top of the original input, showing which parts of the image were used to recognize the object.

At this stage, one can recognize the obvious issue: GRAD-CAM points to different parts of the image. But someone who is blind or low-vision may not be able to perceive the image in the first place. The explanation is useless to them! There's a \textit{perceptual gap} created by the fact the inputs and explanations are both in a particular modality, while the user's perception or mode of sensing isn't.

Users without any sensory disabilities are always able to examine the input to determine if a model is reliable or not \cite{hoiem_diagnosing_2012, bolya_tide_2020, errattahi_automatic_2018}. But disabled users are typically unable to do that without external support \cite{alharbi_misfitting_2024}, which is a major blow to their independence. To make matters worse, the majority of the applications listed above are proprietary, so it is also not possible for external reviewers or auditors to analyze these models either.

Explainability is also crucial for accountability \cite{bellotti_intelligibility_2001, cobbe_reviewable_2021}. If an AI system makes a mistake, who is at fault? The designers of the system for not making it sufficiently robust? The user for using it in scenarios it was unintended for? The annotators further up in the pipeline for inaccurately labelling the training data? The blame game is an easy one to play, but without any explanation for the model's behaviour it's also a fruitless endeavour. And once again, disabled users are unable to make effective use of this facility.

Hence, we ask two questions: 
\begin{itemize}
\item[\textbf{RQ1}:] To what extent does existing work in Explainable AI support or center the needs of users with disabilities?
\item[\textbf{RQ2}:] How can researchers in Accessibility and Explainable AI work together to improve the accessibility and usage of XAI in assistive technologies?
\end{itemize}

To that end, we survey existing literature in Accessibility and XAI to understand where researchers have highlighted such concerns previously. We find only a handful of works have considered this problem, and that there has been almost no work done on integrating Accessible XAI in assistive technology. Following the survey, we offer some suggestions on how can XAI methodology be made more accessible.

Broadly, we offer a provocation to Explainable AI: \textit{How can people participate as shareholders by acknowledging explanations and contest errors in AI outputs when they are incapable of perceiving the ground truth?}

\subsection{Explainability and Accessibility}
\label{sec:Background}

The ways in which XAI fails people with disabilities may not be intuitive to abled people. Before focusing on the two research questions, an accounting of existing XAI methods, the scenarios where they are commonly applied, and how they fail people with disabilities, is needed. 



\subsubsection{Explaining Vision}
\label{sec:introvisualxai}

Vision, both as a modality and as an accessibility problem, is more thoroughly represented than any other. A majority of accessibility work is done for the benefit of BLV people \cite{mack_what_2021}, and a majority of research into human senses in general also focuses on visual perception \cite{hutmacher_why_2019}. An extremely wide body of research exists on Visual XAI, arising mainly from the original CAM paper, such as GradCAM, SmoothCam, ScoreCAM, LayerCAM \cite{selvaraju_grad-cam_2017, omeiza_smooth_2019, jiang_layercam_2021, wang_score-cam_2020}. An alternative are Prototypes \cite{fedele_this_2025, chen_this_2019}, which take parts of the original input directly rather than generating heat maps. Visual models also easily lend themselves to counterfactual explanations \cite{wachter_counterfactual_2017, byrne_counterfactuals_2019}.

BLV users are excluded from almost all of those aforementioned methods, since they can perceive neither the original image, nor the generated heat map or prototype. Existing apps for BLV users tend to rely on uncertainty percentages, which don't 'explain' much \cite{alharbi_misfitting_2024}. That said, there is one method they aren't excluded from: Language.

Visual models can also have language-based explanations \cite{hendricks_generating_2016}, as shown in Figure \ref{fig:VisualExplanationHeader}. Language is versatile and can be expressed in multiple modalities: visually (text), aurally (speech), and tactilely (braille), making it uniquely suited to accessibility contexts. However, they still have drawbacks --- speaking can be slow, while braille displays are specialized devices not available everywhere. Explanations can be unhelpful, focusing on the wrong details, or simply be incorrect \cite{chang_worldscribe_2024, stangl_going_2021, gonzalez_penuela_investigating_2024}.


\subsubsection{Explaining Speech and Language}
\label{sec:introaudioxai}

There is not as much work on the Explainability of speech and language models in general. A wide range of tasks fall under this category: captioning speech, describing or classifying sounds in general, or recognizing and generating sign language.

For speech captioning, feature-based explanations are often used \cite{akman_audio_2024}, alongside breaking down individual phonemes \cite{wu_can_2024}. Confidence scores can also help give users some idea of which parts of the captions are correct or incorrect \cite{kuhn_evaluating_2025}, but there are challenges for comprehensibility in real-world applications \cite{berke_deaf_2017}. 

For audio descriptions, features, spectrogram images (which are difficult for laypeople to parse), and language descriptions \cite{akman_improving_2025} are used. Sign language is typically visual, so there's no perceptual gap for deaf users. However, deaf-blind users still face challenges, and research on tactile sign languages is very sparse.

The perceptual gap in Audio XAI isn't as severe as in Visual XAI. Some existing explanation methods in audio, such as spectrograms, are already visual rather than auditory. Reading text is also much faster than listening to speech, so language descriptions work better. That said, simply explaining isn't the entire goal here: good explanations can teach users how the models work, and very little work has been done in this aspect for audio.

\section{Previous Work}
\label{sec:Previous_Work}


\subsection{Search Methodology}
\label{sec:Methodology}


For \textbf{RQ1}, we focus our search on the ACM Digital Library, where the major venues for disseminating accessibility and human-computer interaction research publish their proceedings \cite{mack_what_2021}. We select four primary venues for accessibility research: ASSETS, TACCESS, CHI, and TOCHI. Works that were published in other venues, were known to the authors prior to the search, or were found in preliminary searches, are listed under 'Other'. We applied the following filters progresssively, and Table \ref{tab:papers_counts}. shows the resulting counts:

\begin{quote}
    \texttt{\textbf{Access}}: accessib* \texttt{OR} disab* \texttt{OR} ``impaired''
\end{quote}

\begin{quote}
    \texttt{\textbf{AI/ML}}: ``machine learning'' \texttt{OR} ``ML'' \texttt{OR} ``computer vision'' \texttt{OR} ``object recognition'' \texttt{OR} ``image captioning'' \texttt{OR} ``automatic speech recognition'' \texttt{OR} ``ASR'' \texttt{OR} ``speech synthesis'' \texttt{OR} ``natural language processing'' \texttt{OR} ``NLP'' \texttt{OR} ``language model''
\end{quote}

\begin{quote}
    \texttt{\textbf{XAI}}: explainab* \texttt{OR} interpretab* \texttt{OR} ``XAI''
\end{quote}

Many of these papers simply use the term "explainable" or "interpretability" in a way unrelated to AI, or mention Explainable AI as a part of the introduction or background only. Hence, we manually verified each result from the final XAI filter. Other terms related to XAI, such as 'uncertainty', 'calibration' or 'trust' returned too many results to verify, and the majority were still unrelated to the problem at hand. So, while there may be some gaps in our review, it's clear that both the Human-Centered AI and the Accessibility communities at large have yet to focus on this intersection.

\begin{table}[h]
\centering
\begin{tabular}{l|cccc}
\hline
\textbf{Venue} & \textbf{Access} & \textbf{AI/ML} & \textbf{XAI} & \textbf{Verified} \\
\hline
ASSETS & 1,097 & 426 & 34 & 3 \\
CHI & 1,083 & 393 & 54 & 1 \\
TACCESS & 261 & 117 & 12 & 1 \\
TOCHI & 422 & 156 & 47 & 0 \\
Other & - & - & - & 5 \\
\hline
Total & 2936 & 1,092 & 147 & 10 \\
\hline
\end{tabular}
\caption{Count of papers per venue based on the above filters.}
\label{tab:papers_counts}
\end{table}

Other venues and workshops that focus on XAI or Accessibility specifically \cite{ehsan_new_2025, sukrut_rao_4th_2025} also do not have any work relevant to the accessibility of XAI.

\subsection{Discussion}
\label{sec:discussion}

These results are unfortunate, but not unexpected. Previous work \cite{suh_fewer_2025, siu_explainable_2025, peixoto_who_2025} has shown only a small fraction of XAI research actually validates their systems through user studies, and the intersection of sensory disabilities and XAI is even more niche.

The closest work to this paper is a similar survey on Accessible Explainable AI by \citet{nwokoye_survey_2024}. Compared to that survey, we focus more more narrowly on the HCI and Accessibility communities, while casting a broader net in terms of \textbf{RQ1} --- we also discuss upstream problems that lead to the need for XAI, or parallel works that help users understand the models without explicitly using XAI methodology. As a position paper, we aim to also situate this problem in the broader canon of human-centered XAI.

At first glance, \cite{diaz-rodriguez_accessible_2020} seems unrelated to accessibility. However, it uses the example of BLV or DHH visitors experiencing cultural heritages as a place where accessible XAI may be needed. \citet{wolf_designing_2020} similarly notes that all explanations should be tailored to specific audiences and use appropriate interfaces. \citet{peixoto_who_2025} both surveys XAI research and sets forth similar guidelines.

Both \citet{alharbi_misfitting_2024} and \citet{fernando_image_2025} approach the problem of trust and verifiability for BLV users of accessibility software qualitatively. The interviewed participants highlight multiple problems with existing applications: that they only express uncertainty through a single 'confidence' score value, and it takes a long time or a lot of effort to interact with responses from the app. These issues boil down to the fact that XAI methods are not designed for users like them. \citet{khan_tell_2025} focuses specifically on the issues brought up in the \textit{reviews} of apps used by BLV users. While most reviews are positive, some reviews also mention similar problems with accuracy and the consequences of an incorrect output. \citet{hong_understanding_2024} go into the details of how blind users handle errors, and note how unsuitable existing XAI techniques are for blind users.

ImageExplorer \cite{lee_imageexplorer_2022} uses scene graphs to break down images into components for easier interrogation. This lets users interact with the image and understand if an output from a model is accurate or not. While not strictly using an XAI method, this approach is similar to how abled users might interact with prototypical or heat map explanations. 

Finally, \citet{peixoto_exploring_2025} directly engages with the problem of XAI in Accessibility tools. They discuss how BLV users can interact with audio descriptions as XAI for automatic video captioning, and also survey existing work in sound-based XAI.

All of the above focus on visually-disabled users or visual models. What about audio and speech? As mentioned in Section \ref{sec:introaudioxai}, audio suffers from less of a perceptual gap. However, XAI is still both underutilized and understudied in this area, and a few works such as \cite{kafle_artificial_2020, bragg_fate_2021} have called attention to it.

\section{Making XAI Accessible}
\label{sec:makingxai}

In this section, we focus on \textbf{RQ2} and offer some suggestions to XAI, Accessibility and even Visualization researchers to help solve the problems described so far. This is not an exhaustive list, and is phrased as questions and suggestions for future research directions that would improve the trustworthiness of assistive technology that uses AI, allowing disabled users to maintain a greater degree of independence.

\subsection{What is Verifiable and what is not?}
\label{sec:verifiable}

As Figure \ref{fig:VisualExplanationHeader} shows, a language description may have some components that can be verified using other senses (touch, hearing or smell) while other components may be hard or impossible to verify (background details or colour). The 'verifiability' of a component can also depend on the context. E.g., the background of an image inside one's own house is verifiable, but not of an outdoors image. 

In the near term, future work could consider integrating the idea of verifiability into multimodal models (Sections \ref{sec:languagedesc} and \ref{sec:multimodal}). In the long run, we need to understand what people with sensory disabilities consider 'verifiable', and how to personalize such systems to account for varying contexts, use cases, and individual needs.

\subsection{Where do Language Descriptions fall short, and how can they be improved?} 
\label{sec:languagedesc}

Existing language descriptions rely on standard visual captioning models, which are annotated by abled people. When used in accessibility applications downstream, they often veer away from disabled users' needs \cite{garg_its_2025}. \citet{gurari_vizwiz_2018}'s VizWiz dataset takes a step in the right direction, involving disabled people in the data collection part. However, the labelling and captioning afterwards were still done by abled people.

In the near term, researchers should aim to simply create more accurate vision-language models that give verifiable (Section \ref{sec:verifiable}) descriptions, or emulate the perceptual experience of disabled users better. In the long term, people with disabilities should be included at every step of the process of training models, which means we must design new interfaces and procedures for them to interact with training data and model evaluations in detail.

\subsection{Can XAI Visualizations be made more Accessible?}

Feature-based explanations are typically visualized as graphs or charts \cite{bhati_survey_2025}, which can be effectively described \cite{siu_supporting_2022}. Heat-map-based visual explanations pose a greater challenge, but could be combined with methods like \cite{lee_imageexplorer_2022} and \cite{biggs_audio_2018} to create accessible explanations.

Many methods of visualization are now more accessible \cite{hong_understanding_2024}. In future research, we encourage Visualization, XAI and Accessibility researchers to collaborate and implement these advances into existing assistive software or into the training process of machine learning models.

\subsection{Can better Explanations lead to better Teaching?}

Teachable AI describes systems that are designed from the ground up to be personalizable to the needs of a very small group or even a single individual \cite{kacorri_teachable_2017, kacorri_people_2017, morrison_understanding_2023, wen_find_2024}. 

Since Teachable AI systems are not infallible, future work in this area should consider integrating XAI into them. That would both help disabled users better understand the cause behind those failures \cite{kaur_interpreting_2020} and 'teach' the systems better, while also improving education and trust about such AI systems in general.

\subsection{Can Explaining Multimodal \textit{inputs} improve Trust and Verifiability?}
\label{sec:multimodal}

People with sensory disabilities rely on multiple other senses. For instance, A DHH person might feel vibrations on the floor, and intuit someone's running around nearby even if they cannot see the other person. This principle can be extended to modern multimodal models as well. Even tactile models \cite{yang_binding_2024} could be used here, basing outputs off vibrations, impact, and the feel of different materials.

In future work, multimodal inputs can be a pathway to improving verifiability (Section \ref{sec:verifiable}) in assistive technologies.

\section{Conclusion}




Many modern assistive apps and tools now use machine learning models. Since these models are not always reliable, it is important that users are able to understand the processes or verify the outputs. Explainable AI is the primary tool for solving this problem. However, due to a perceptual gap, users with sensory disabilities are excluded from most XAI methods.

Hence, we propose some general research directions focused on two key points --- that explanations and outputs should be designed in such a way as to be verifiable by disabled users, and that disabled users should be able to participate at all steps of the training process of the models they will use. We hope this paper encourages researchers to look deeper at the largely understudied intersection of Accessibility and Explainable AI, to improve trust and reliability, and allow people with disabilities to empower their use of assistive technologies and maintain independent lives.


\bibliographystyle{ACM-Reference-Format}
\bibliography{sample-base}

@String{Computing = "Computing" }

@String{Computer = "{IEEE} Computer" }

@String{Springer = "Springer-Verlag" }

@inproceedings{kaur_interpreting_2020,
	address = {New York, NY, USA},
	series = {{CHI} '20},
	title = {Interpreting {Interpretability}: {Understanding} {Data} {Scientists}' {Use} of {Interpretability} {Tools} for {Machine} {Learning}},
	isbn = {978-1-4503-6708-0},
	shorttitle = {Interpreting {Interpretability}},
	url = {https://doi.org/10.1145/3313831.3376219},
	doi = {10.1145/3313831.3376219},
	urldate = {2025-09-20},
	booktitle = {Proceedings of the 2020 {CHI} {Conference} on {Human} {Factors} in {Computing} {Systems}},
	publisher = {Association for Computing Machinery},
	author = {Kaur, Harmanpreet and Nori, Harsha and Jenkins, Samuel and Caruana, Rich and Wallach, Hanna and Wortman Vaughan, Jennifer},
	month = apr,
	year = {2020},
	pages = {1--14},
}

@inproceedings{ehsan_new_2025,
	address = {New York, NY, USA},
	series = {{CHI} {EA} '25},
	title = {New {Frontiers} of {Human}-centered {Explainable} {AI} ({HCXAI}): {Participatory} {Civic} {AI}, {Benchmarking} {LLMs}, {XAI} {Hallucinations}, and {Responsible} {AI} {Audits}},
	isbn = {979-8-4007-1395-8},
	shorttitle = {New {Frontiers} of {Human}-centered {Explainable} {AI} ({HCXAI})},
	url = {https://dl.acm.org/doi/10.1145/3706599.3706713},
	doi = {10.1145/3706599.3706713},
	urldate = {2026-02-01},
	booktitle = {Proceedings of the {Extended} {Abstracts} of the {CHI} {Conference} on {Human} {Factors} in {Computing} {Systems}},
	publisher = {Association for Computing Machinery},
	author = {Ehsan, Upol and Watkins, Elizabeth A and Wintersberger, Philipp and Manger, Carina and Hubig, Nina and Savage, Saiph and Weisz, Justin D. and Riener, Andreas},
	month = apr,
	year = {2025},
	pages = {1--6},
}

@misc{sukrut_rao_4th_2025,
	title = {The 4th {Explainable} {AI} for {Computer} {Vision} ({XAI4CV}) {Workshop}},
	url = {https://xai4cv.github.io/workshop_cvpr25},
	urldate = {2026-02-01},
    year = {2025},
	author = {{Sukrut Rao} and {Indu Panigrahi} and Kim, Sunnie S. Y. and {Vikram V. Ramaswamy} and {Rajat Sahay} and {Avinab Saha} and {Dahye Kim} and {Miguel-Ángel Fernández-Torres} and {Lenka Tětková} and {Teresa Dorszewski} and {Bartlomiej Sobieski} and {Marina Gavrilova} and {Yuhui Zhang} and {Pushkar Shukla}},
}

@article{kacorri_teachable_2017,
	title = {Teachable machines for accessibility},
	issn = {1558-2337},
	url = {https://dl.acm.org/doi/10.1145/3167902.3167904},
	doi = {10.1145/3167902.3167904},
	number = {119},
	urldate = {2026-02-01},
	journal = {SIGACCESS Access. Comput.},
	author = {Kacorri, Hernisa},
	month = nov,
	year = {2017},
	pages = {10--18},
}

@inproceedings{morrison_understanding_2023,
	address = {New York, NY, USA},
	series = {{ASSETS} '23},
	title = {Understanding {Personalized} {Accessibility} through {Teachable} {AI}: {Designing} and {Evaluating} {Find} {My} {Things} for {People} who are {Blind} or {Low} {Vision}},
	isbn = {979-8-4007-0220-4},
	shorttitle = {Understanding {Personalized} {Accessibility} through {Teachable} {AI}},
	url = {https://dl.acm.org/doi/10.1145/3597638.3608395},
	doi = {10.1145/3597638.3608395},
	urldate = {2026-02-01},
	booktitle = {Proceedings of the 25th {International} {ACM} {SIGACCESS} {Conference} on {Computers} and {Accessibility}},
	publisher = {Association for Computing Machinery},
	author = {Morrison, Cecily and Grayson, Martin and Marques, Rita Faia and Massiceti, Daniela and Longden, Camilla and Wen, Linda and Cutrell, Edward},
	month = oct,
	year = {2023},
	pages = {1--12},
}

@inproceedings{wen_find_2024,
	address = {New York, NY, USA},
	series = {{CHI} {EA} '24},
	title = {Find {My} {Things}: {Personalized} {Accessibility} through {Teachable} {AI} for {People} who are {Blind} or {Low} {Vision}},
	isbn = {979-8-4007-0331-7},
	shorttitle = {Find {My} {Things}},
	url = {https://dl.acm.org/doi/10.1145/3613905.3648641},
	doi = {10.1145/3613905.3648641},
	urldate = {2026-02-01},
	booktitle = {Extended {Abstracts} of the {CHI} {Conference} on {Human} {Factors} in {Computing} {Systems}},
	publisher = {Association for Computing Machinery},
	author = {Wen, Linda Yilin and Morrison, Cecily and Grayson, Martin and Marques, Rita Faia and Massiceti, Daniela and Longden, Camilla and Cutrell, Edward},
	month = may,
	year = {2024},
	pages = {1--6},
}

@inproceedings{kacorri_people_2017,
	address = {New York, NY, USA},
	series = {{CHI} '17},
	title = {People with {Visual} {Impairment} {Training} {Personal} {Object} {Recognizers}: {Feasibility} and {Challenges}},
	isbn = {978-1-4503-4655-9},
	shorttitle = {People with {Visual} {Impairment} {Training} {Personal} {Object} {Recognizers}},
	url = {https://dl.acm.org/doi/10.1145/3025453.3025899},
	doi = {10.1145/3025453.3025899},
	urldate = {2026-02-01},
	booktitle = {Proceedings of the 2017 {CHI} {Conference} on {Human} {Factors} in {Computing} {Systems}},
	publisher = {Association for Computing Machinery},
	author = {Kacorri, Hernisa and Kitani, Kris M. and Bigham, Jeffrey P. and Asakawa, Chieko},
	month = may,
	year = {2017},
	pages = {5839--5849},
}

@inproceedings{chang_worldscribe_2024,
	address = {New York, NY, USA},
	series = {{UIST} '24},
	title = {{WorldScribe}: {Towards} {Context}-{Aware} {Live} {Visual} {Descriptions}},
	isbn = {979-8-4007-0628-8},
	shorttitle = {{WorldScribe}},
	url = {https://dl.acm.org/doi/10.1145/3654777.3676375},
	doi = {10.1145/3654777.3676375},
	urldate = {2026-02-01},
	booktitle = {Proceedings of the 37th {Annual} {ACM} {Symposium} on {User} {Interface} {Software} and {Technology}},
	publisher = {Association for Computing Machinery},
	author = {Chang, Ruei-Che and Liu, Yuxuan and Guo, Anhong},
	month = oct,
	year = {2024},
	pages = {1--18},
}

@inproceedings{siu_supporting_2022,
	address = {New York, NY, USA},
	series = {{CHI} '22},
	title = {Supporting {Accessible} {Data} {Visualization} {Through} {Audio} {Data} {Narratives}},
	isbn = {978-1-4503-9157-3},
	url = {https://dl.acm.org/doi/10.1145/3491102.3517678},
	doi = {10.1145/3491102.3517678},
	urldate = {2026-02-01},
	booktitle = {Proceedings of the 2022 {CHI} {Conference} on {Human} {Factors} in {Computing} {Systems}},
	publisher = {Association for Computing Machinery},
	author = {Siu, Alexa and S-H Kim, Gene and O'Modhrain, Sile and Follmer, Sean},
	month = apr,
	year = {2022},
	pages = {1--19},
}

@inproceedings{gonzalez_penuela_investigating_2024,
	address = {New York, NY, USA},
	series = {{CHI} '24},
	title = {Investigating {Use} {Cases} of {AI}-{Powered} {Scene} {Description} {Applications} for {Blind} and {Low} {Vision} {People}},
	isbn = {979-8-4007-0330-0},
	url = {https://dl.acm.org/doi/10.1145/3613904.3642211},
	doi = {10.1145/3613904.3642211},
	urldate = {2026-02-01},
	booktitle = {Proceedings of the 2024 {CHI} {Conference} on {Human} {Factors} in {Computing} {Systems}},
	publisher = {Association for Computing Machinery},
	author = {Gonzalez Penuela, Ricardo E and Collins, Jazmin and Bennett, Cynthia and Azenkot, Shiri},
	month = may,
	year = {2024},
	pages = {1--21},
}

@inproceedings{stangl_going_2021,
	address = {New York, NY, USA},
	series = {{ASSETS} '21},
	title = {Going {Beyond} {One}-{Size}-{Fits}-{All} {Image} {Descriptions} to {Satisfy} the {Information} {Wants} of {People} {Who} are {Blind} or {Have} {Low} {Vision}},
	isbn = {978-1-4503-8306-6},
	url = {https://dl.acm.org/doi/10.1145/3441852.3471233},
	doi = {10.1145/3441852.3471233},
	urldate = {2026-02-01},
	booktitle = {Proceedings of the 23rd {International} {ACM} {SIGACCESS} {Conference} on {Computers} and {Accessibility}},
	publisher = {Association for Computing Machinery},
	author = {Stangl, Abigale and Verma, Nitin and Fleischmann, Kenneth R. and Morris, Meredith Ringel and Gurari, Danna},
	month = oct,
	year = {2021},
	pages = {1--15},
}

@inproceedings{yang_binding_2024,
	title = {Binding {Touch} to {Everything}: {Learning} {Unified} {Multimodal} {Tactile} {Representations}},
	shorttitle = {Binding {Touch} to {Everything}},
	url = {https://openaccess.thecvf.com/content/CVPR2024/html/Yang_Binding_Touch_to_Everything_Learning_Unified_Multimodal_Tactile_Representations_CVPR_2024_paper.html},
	language = {en},
	urldate = {2026-01-23},
	author = {Yang, Fengyu and Feng, Chao and Chen, Ziyang and Park, Hyoungseob and Wang, Daniel and Dou, Yiming and Zeng, Ziyao and Chen, Xien and Gangopadhyay, Rit and Owens, Andrew and Wong, Alex},
	year = {2024},
	pages = {26340--26353},
}

@article{peixoto_exploring_2025,
	title = {Exploring {Accessible} {Explainable} {AI}: {Promising} {Avenues}},
	shorttitle = {Exploring {Accessible} {Explainable} {AI}},
	url = {https://scholarworks.calstate.edu/downloads/2f75rj705#page=358},
	urldate = {2026-01-23},
	journal = {The Journal on Technology and Persons with Disabilities},
	author = {Peixoto, Maria JP and Nwokoye, Chukwunonso Henry and Pandey, Akriti and Zaman, Ahsan and Lewis, Peter R.},
	year = {2025},
	pages = {350},
}

@misc{garg_its_2025,
	title = {"{It}'s trained by non-disabled people": {Evaluating} {How} {Image} {Quality} {Affects} {Product} {Captioning} with {VLMs}},
	shorttitle = {"{It}'s trained by non-disabled people"},
	url = {http://arxiv.org/abs/2511.08917},
	doi = {10.48550/arXiv.2511.08917},
	urldate = {2026-01-23},
	publisher = {arXiv},
	author = {Garg, Kapil and Tang, Xinru and Heo, Jimin and Morgan, Dwayne R. and Gergle, Darren and Sudderth, Erik B. and Piper, Anne Marie},
	month = nov,
	year = {2025},
	note = {arXiv:2511.08917 [cs]},
}

@article{kafle_artificial_2020,
	title = {Artificial intelligence fairness in the context of accessibility research on intelligent systems for people who are deaf or hard of hearing},
	issn = {1558-2337},
	url = {https://dl.acm.org/doi/10.1145/3386296.3386300},
	doi = {10.1145/3386296.3386300},
	number = {125},
	urldate = {2026-01-23},
	journal = {SIGACCESS Access. Comput.},
	author = {Kafle, Sushant and Glasser, Abraham and Al-khazraji, Sedeeq and Berke, Larwan and Seita, Matthew and Huenerfauth, Matt},
	month = mar,
	year = {2020},
	pages = {4:1},
}

@article{bragg_fate_2021,
	title = {The {FATE} {Landscape} of {Sign} {Language} {AI} {Datasets}: {An} {Interdisciplinary} {Perspective}},
	volume = {14},
	issn = {1936-7228},
	shorttitle = {The {FATE} {Landscape} of {Sign} {Language} {AI} {Datasets}},
	url = {https://dl.acm.org/doi/10.1145/3436996},
	doi = {10.1145/3436996},
	number = {2},
	urldate = {2025-09-20},
	journal = {ACM Trans. Access. Comput.},
	author = {Bragg, Danielle and Caselli, Naomi and Hochgesang, Julie A. and Huenerfauth, Matt and Katz-Hernandez, Leah and Koller, Oscar and Kushalnagar, Raja and Vogler, Christian and Ladner, Richard E.},
	month = jul,
	year = {2021},
	pages = {7:1--7:45},
}

@article{fernando_image_2025,
	title = {Image {Recognition} {Tools} for {Blind} and {Visually} {Impaired} {Users}: {An} {Emphasis} on the {Design} {Considerations}},
	volume = {18},
	issn = {1936-7228},
	shorttitle = {Image {Recognition} {Tools} for {Blind} and {Visually} {Impaired} {Users}},
	url = {https://dl.acm.org/doi/10.1145/3702208},
	doi = {10.1145/3702208},
	number = {1},
	urldate = {2025-09-20},
	journal = {ACM Trans. Access. Comput.},
	author = {Fernando, Sandra and Ndukwe, Chiemela and Virdee, Bal and Djemai, Ramzi},
	month = jan,
	year = {2025},
	pages = {1:1--1:21},
}

@inproceedings{lee_imageexplorer_2022,
	address = {New York, NY, USA},
	series = {{CHI} '22},
	title = {{ImageExplorer}: {Multi}-{Layered} {Touch} {Exploration} to {Encourage} {Skepticism} {Towards} {Imperfect} {AI}-{Generated} {Image} {Captions}},
	isbn = {978-1-4503-9157-3},
	shorttitle = {{ImageExplorer}},
	url = {https://doi.org/10.1145/3491102.3501966},
	doi = {10.1145/3491102.3501966},
	urldate = {2025-09-20},
	booktitle = {Proceedings of the 2022 {CHI} {Conference} on {Human} {Factors} in {Computing} {Systems}},
	publisher = {Association for Computing Machinery},
	author = {Lee, Jaewook and Herskovitz, Jaylin and Peng, Yi-Hao and Guo, Anhong},
	month = apr,
	year = {2022},
	pages = {1--15},
}

@inproceedings{khan_tell_2025,
	address = {Denver Colorado USA},
	title = {"{Tell} {Me} {What} {That} {Is}": {Examining} {Blind} and {Low} {Vision} {Individuals}’ {Reviews} on {Assistive} {Smartphone} {Applications}},
	isbn = {979-8-4007-0676-9},
	shorttitle = {"{Tell} {Me} {What} {That} {Is}"},
	url = {https://dl.acm.org/doi/10.1145/3663547.3759699},
	doi = {10.1145/3663547.3759699},
	language = {en},
	urldate = {2025-10-24},
	booktitle = {Proceedings of the 27th {International} {ACM} {SIGACCESS} {Conference} on {Computers} and {Accessibility}},
	publisher = {ACM},
	author = {Khan, Waqar Hassan and Buo, Isaac and Middel, Ariane},
	month = oct,
	year = {2025},
	pages = {1--6},
}

@inproceedings{diaz-rodriguez_accessible_2020,
	address = {New York, NY, USA},
	series = {{UMAP} '20 {Adjunct}},
	title = {Accessible {Cultural} {Heritage} through {Explainable} {Artificial} {Intelligence}},
	isbn = {978-1-4503-7950-2},
	url = {https://doi.org/10.1145/3386392.3399276},
	doi = {10.1145/3386392.3399276},
	urldate = {2025-10-23},
	booktitle = {Adjunct {Publication} of the 28th {ACM} {Conference} on {User} {Modeling}, {Adaptation} and {Personalization}},
	publisher = {Association for Computing Machinery},
	author = {Díaz-Rodríguez, Natalia and Pisoni, Galena},
	month = jul,
	year = {2020},
	pages = {317--324},
}

@article{wolf_designing_2020,
	title = {Designing accessible, explainable {AI} ({XAI}) experiences},
	issn = {1558-2337},
	url = {https://doi.org/10.1145/3386296.3386302},
	doi = {10.1145/3386296.3386302},
	number = {125},
	urldate = {2025-10-19},
	journal = {SIGACCESS Access. Comput.},
	author = {Wolf, Christine T. and Ringland, Kathryn E.},
	month = mar,
	year = {2020},
	pages = {6:1},
}

@article{akman_improving_2025,
	title = {Improving {Audio} {Explanations} {Using} {Audio} {Language} {Models}},
	volume = {32},
	issn = {1558-2361},
	url = {https://ieeexplore.ieee.org/abstract/document/10847866},
	doi = {10.1109/LSP.2025.3532218},
	urldate = {2026-01-22},
	journal = {IEEE Signal Processing Letters},
	author = {Akman, Alican and Sun, Qiyang and Schuller, Björn W.},
	year = {2025},
	pages = {741--745},
}

@inproceedings{berke_deaf_2017,
	address = {New York, NY, USA},
	series = {{ASSETS} '17},
	title = {Deaf and {Hard}-of-{Hearing} {Perspectives} on {Imperfect} {Automatic} {Speech} {Recognition} for {Captioning} {One}-on-{One} {Meetings}},
	isbn = {978-1-4503-4926-0},
	url = {https://dl.acm.org/doi/10.1145/3132525.3132541},
	doi = {10.1145/3132525.3132541},
	urldate = {2026-01-22},
	booktitle = {Proceedings of the 19th {International} {ACM} {SIGACCESS} {Conference} on {Computers} and {Accessibility}},
	publisher = {Association for Computing Machinery},
	author = {Berke, Larwan and Caulfield, Christopher and Huenerfauth, Matt},
	month = oct,
	year = {2017},
	pages = {155--164},
}

@inproceedings{wu_can_2024,
	title = {Can {We} {Trust} {Explainable} {AI} {Methods} on {ASR}? {An} {Evaluation} on {Phoneme} {Recognition}},
	shorttitle = {Can {We} {Trust} {Explainable} {AI} {Methods} on {ASR}?},
	url = {https://ieeexplore.ieee.org/abstract/document/10445989},
	doi = {10.1109/ICASSP48485.2024.10445989},
	urldate = {2026-01-22},
	booktitle = {{ICASSP} 2024 - 2024 {IEEE} {International} {Conference} on {Acoustics}, {Speech} and {Signal} {Processing} ({ICASSP})},
	author = {Wu, Xiaoliang and Bell, Peter and Rajan, Ajitha},
	month = apr,
	year = {2024},
	note = {ISSN: 2379-190X},
	pages = {10296--10300},
}

@article{akman_audio_2024,
	title = {Audio {Explainable} {Artificial} {Intelligence}: {A} {Review}},
	volume = {3},
	shorttitle = {Audio {Explainable} {Artificial} {Intelligence}},
	url = {https://spj.science.org/doi/full/10.34133/icomputing.0074},
	doi = {10.34133/icomputing.0074},
	urldate = {2026-01-22},
	journal = {Intelligent Computing},
	author = {Akman, Alican and Schuller, Björn W.},
	month = jan,
	year = {2024},
	note = {Publisher: American Association for the Advancement of Science},
	pages = {0074},
}

@article{hutmacher_why_2019,
	title = {Why {Is} {There} {So} {Much} {More} {Research} on {Vision} {Than} on {Any} {Other} {Sensory} {Modality}?},
	volume = {10},
	issn = {1664-1078},
	url = {https://pmc.ncbi.nlm.nih.gov/articles/PMC6787282/},
	doi = {10.3389/fpsyg.2019.02246},
	urldate = {2026-01-22},
	journal = {Frontiers in Psychology},
	author = {Hutmacher, Fabian},
	month = oct,
	year = {2019},
	pmid = {31636589},
	pmcid = {PMC6787282},
	pages = {2246},
}

@misc{nwokoye_survey_2024,
	title = {A {Survey} of {Accessible} {Explainable} {Artificial} {Intelligence} {Research}},
	url = {http://arxiv.org/abs/2407.17484},
	doi = {10.48550/arXiv.2407.17484},
	urldate = {2024-11-16},
	publisher = {arXiv},
	author = {Nwokoye, Chukwunonso Henry and Peixoto, Maria J. P. and Pandey, Akriti and Pardy, Lauren and Sukhai, Mahadeo and Lewis, Peter R.},
	month = jul,
	year = {2024},
	note = {arXiv:2407.17484 
version: 1},
}

@inproceedings{alharbi_misfitting_2024,
	address = {New York, NY, USA},
	series = {{ASSETS} '24},
	title = {Misfitting {With} {AI}: {How} {Blind} {People} {Verify} and {Contest} {AI} {Errors}},
	isbn = {979-8-4007-0677-6},
	shorttitle = {Misfitting {With} {AI}},
	url = {https://dl.acm.org/doi/10.1145/3663548.3675659},
	doi = {10.1145/3663548.3675659},
	urldate = {2025-03-14},
	booktitle = {Proceedings of the 26th {International} {ACM} {SIGACCESS} {Conference} on {Computers} and {Accessibility}},
	publisher = {Association for Computing Machinery},
	author = {Alharbi, Rahaf and Lor, Pa and Herskovitz, Jaylin and Schoenebeck, Sarita and Brewer, Robin N.},
	month = oct,
	year = {2024},
	pages = {1--17},
}

@inproceedings{lundberg_unified_2017,
	address = {Red Hook, NY, USA},
	series = {{NIPS}'17},
	title = {A unified approach to interpreting model predictions},
	isbn = {978-1-5108-6096-4},
	urldate = {2025-08-16},
	booktitle = {Proceedings of the 31st {International} {Conference} on {Neural} {Information} {Processing} {Systems}},
	publisher = {Curran Associates Inc.},
	author = {Lundberg, Scott M. and Lee, Su-In},
	month = dec,
	year = {2017},
	pages = {4768--4777},
}

@inproceedings{ribeiro_why_2016,
	address = {New York, NY, USA},
	series = {{KDD} '16},
	title = {"{Why} {Should} {I} {Trust} {You}?": {Explaining} the {Predictions} of {Any} {Classifier}},
	isbn = {978-1-4503-4232-2},
	shorttitle = {"{Why} {Should} {I} {Trust} {You}?},
	url = {https://dl.acm.org/doi/10.1145/2939672.2939778},
	doi = {10.1145/2939672.2939778},
	urldate = {2025-08-16},
	booktitle = {Proceedings of the 22nd {ACM} {SIGKDD} {International} {Conference} on {Knowledge} {Discovery} and {Data} {Mining}},
	publisher = {Association for Computing Machinery},
	author = {Ribeiro, Marco Tulio and Singh, Sameer and Guestrin, Carlos},
	month = aug,
	year = {2016},
	pages = {1135--1144},
}

@inproceedings{zhou_learning_2016,
	title = {Learning {Deep} {Features} for {Discriminative} {Localization}},
	url = {https://openaccess.thecvf.com/content_cvpr_2016/html/Zhou_Learning_Deep_Features_CVPR_2016_paper.html},
	urldate = {2025-08-01},
	author = {Zhou, Bolei and Khosla, Aditya and Lapedriza, Agata and Oliva, Aude and Torralba, Antonio},
	year = {2016},
	pages = {2921--2929},
}

@inproceedings{selvaraju_grad-cam_2017,
	title = {Grad-{CAM}: {Visual} {Explanations} from {Deep} {Networks} via {Gradient}-{Based} {Localization}},
	shorttitle = {Grad-{CAM}},
	url = {https://ieeexplore.ieee.org/document/8237336},
	doi = {10.1109/ICCV.2017.74},
	urldate = {2025-03-14},
	booktitle = {2017 {IEEE} {International} {Conference} on {Computer} {Vision} ({ICCV})},
	author = {Selvaraju, Ramprasaath R. and Cogswell, Michael and Das, Abhishek and Vedantam, Ramakrishna and Parikh, Devi and Batra, Dhruv},
	month = oct,
	year = {2017},
	note = {ISSN: 2380-7504},
	pages = {618--626},
}

@inproceedings{biggs_audio_2018,
	address = {Houghton, USA},
	title = {The audio game laboratory: {Building} maps from games},
	copyright = {cc\_by\_nc\_4},
	shorttitle = {The audio game laboratory},
	url = {http://icad2018.icad.org/wp-content/uploads/2018/06/ICAD2018_paper_10.pdf},
	language = {en},
	urldate = {2026-01-23},
	author = {Biggs, Brandon and Yusim, Lena and Coppin, Peter},
	month = jul,
	year = {2018},
}

@article{bhati_survey_2025,
	title = {A {Survey} of {Post}-{Hoc} {XAI} {Methods} {From} a {Visualization} {Perspective}: {Challenges} and {Opportunities}},
	volume = {13},
	issn = {2169-3536},
	shorttitle = {A {Survey} of {Post}-{Hoc} {XAI} {Methods} {From} a {Visualization} {Perspective}},
	url = {https://ieeexplore.ieee.org/abstract/document/11039632},
	doi = {10.1109/ACCESS.2025.3581136},
	urldate = {2026-01-23},
	journal = {IEEE Access},
	author = {Bhati, Deepshikha and Amiruzzaman, MD and Zhao, Ye and Guercio, Angela and Le, Tram},
	year = {2025},
	pages = {120785--120806},
}

@inproceedings{hong_understanding_2024,
	address = {St. John's NL Canada},
	title = {Understanding {How} {Blind} {Users} {Handle} {Object} {Recognition} {Errors}: {Strategies} and {Challenges}},
	isbn = {979-8-4007-0677-6},
	shorttitle = {Understanding {How} {Blind} {Users} {Handle} {Object} {Recognition} {Errors}},
	url = {https://dl.acm.org/doi/10.1145/3663548.3675635},
	doi = {10.1145/3663548.3675635},
	language = {en},
	urldate = {2025-04-10},
	booktitle = {The 26th {International} {ACM} {SIGACCESS} {Conference} on {Computers} and {Accessibility}},
	publisher = {ACM},
	author = {Hong, Jonggi and Kacorri, Hernisa},
	month = oct,
	year = {2024},
	pages = {1--15},
}

@inproceedings{hoiem_diagnosing_2012,
	address = {Berlin, Heidelberg},
	title = {Diagnosing {Error} in {Object} {Detectors}},
	isbn = {978-3-642-33712-3},
	doi = {10.1007/978-3-642-33712-3_25},
	language = {en},
	booktitle = {Computer {Vision} – {ECCV} 2012},
	publisher = {Springer},
	author = {Hoiem, Derek and Chodpathumwan, Yodsawalai and Dai, Qieyun},
	editor = {Fitzgibbon, Andrew and Lazebnik, Svetlana and Perona, Pietro and Sato, Yoichi and Schmid, Cordelia},
	year = {2012},
	pages = {340--353},
}

@inproceedings{bolya_tide_2020,
	address = {Cham},
	title = {{TIDE}: {A} {General} {Toolbox} for {Identifying} {Object} {Detection} {Errors}},
	isbn = {978-3-030-58580-8},
	shorttitle = {{TIDE}},
	doi = {10.1007/978-3-030-58580-8_33},
	language = {en},
	booktitle = {Computer {Vision} – {ECCV} 2020},
	publisher = {Springer International Publishing},
	author = {Bolya, Daniel and Foley, Sean and Hays, James and Hoffman, Judy},
	editor = {Vedaldi, Andrea and Bischof, Horst and Brox, Thomas and Frahm, Jan-Michael},
	year = {2020},
	pages = {558--573},
}

@article{errattahi_automatic_2018,
	series = {1st {International} {Conference} on {Natural} {Language} and {Speech} {Processing}},
	title = {Automatic {Speech} {Recognition} {Errors} {Detection} and {Correction}: {A} {Review}},
	volume = {128},
	issn = {1877-0509},
	shorttitle = {Automatic {Speech} {Recognition} {Errors} {Detection} and {Correction}},
	url = {https://www.sciencedirect.com/science/article/pii/S1877050918302187},
	doi = {10.1016/j.procs.2018.03.005},
	urldate = {2025-08-01},
	journal = {Procedia Computer Science},
	author = {Errattahi, Rahhal and El Hannani, Asmaa and Ouahmane, Hassan},
	month = jan,
	year = {2018},
	pages = {32--37},
}

@article{bellotti_intelligibility_2001,
	title = {Intelligibility and {Accountability}: {Human} {Considerations} in {Context}-{Aware} {Systems}},
	copyright = {Copyright Taylor and Francis Group, LLC},
	shorttitle = {Intelligibility and {Accountability}},
	url = {https://www.tandfonline.com/doi/abs/10.1207/S15327051HCI16234_05},
	doi = {10.1207/S15327051HCI16234_05},
	language = {EN},
	urldate = {2025-09-12},
	journal = {Human–Computer Interaction},
	author = {Bellotti, Victoria and Edwards, Keith},
	month = dec,
	year = {2001},
	note = {Publisher: Lawrence Erlbaum Associates, Inc.},
}

@inproceedings{cobbe_reviewable_2021,
	address = {Virtual Event Canada},
	title = {Reviewable {Automated} {Decision}-{Making}: {A} {Framework} for {Accountable} {Algorithmic} {Systems}},
	isbn = {978-1-4503-8309-7},
	shorttitle = {Reviewable {Automated} {Decision}-{Making}},
	url = {https://dl.acm.org/doi/10.1145/3442188.3445921},
	doi = {10.1145/3442188.3445921},
	language = {en},
	urldate = {2025-09-12},
	booktitle = {Proceedings of the 2021 {ACM} {Conference} on {Fairness}, {Accountability}, and {Transparency}},
	publisher = {ACM},
	author = {Cobbe, Jennifer and Lee, Michelle Seng Ah and Singh, Jatinder},
	month = mar,
	year = {2021},
	pages = {598--609},
}

\end{document}